\documentclass[aps,
superscriptaddress,amsmath,amssymb,showpacs,twocolumn]{revtex4-1}

\usepackage{graphicx}
\usepackage{hyperref}
\usepackage{epsfig}
\usepackage{color}
\usepackage{psfrag}
\usepackage{float}
\usepackage{bbold}
\usepackage{tikz}

\newcommand{\ket}[1] {\left|#1\right\rangle}

\newcommand{\rem}[1]{}

\newcommand{\refe}[1]{~(\ref{#1})}
\newcommand{\Eqref}[1]{Eq.~(\ref{#1})}

\newcommand{\gammaEM}{\gamma}  
 
\begin{document}

\title{Washing out of the  $0$-$\pi$ transition in Josephson junctions}

\author{R. Avriller} 
\affiliation{Univ.~Bordeaux, LOMA, UMR 5798, F-33400 Talence, France \\
CNRS, LOMA, UMR 5798, F-33400 Talence, France}

\author{F. Pistolesi} 
\affiliation{Univ.~Bordeaux, LOMA, UMR 5798, F-33400 Talence, France \\
CNRS, LOMA, UMR 5798, F-33400 Talence, France}

\date{\today}

\begin{abstract}
We consider a Josephson junction formed by a quantum 
dot connected to two bulk superconductors in presence of 
Coulomb interaction and coupling to both an electromagnetic environment
and a finite density of electronic quasi-particles.
In the limit of large superconducting gap we obtain a Born-Markov description 
of the system dynamics.
We calculate the current-phase relation and we find that the experimentally 
unavoidable presence of quasi-particles can dramatically modify the 
$0$-$\pi$ standard transition picture.
We show that photon-assisted quasi-particles absorption 
allows the dynamic switching from the $0$- to the $\pi$-state and vice-versa,
washing out the $0$-$\pi$ transition predicted by purely thermodynamic arguments.
\end{abstract}


\maketitle

%
%
%
%
\textit{Introduction.---}
The Josephson junction is a fundamental element of superconducting quantum nano-electronics, 
with a wide spectrum of applications ranging from quantum information to medical imagery.
Such a junction can be formed by contacting two superconductors by a large variety of nano-structures
\cite{kasumov_supercurrents_1999, steinbach_direct_2001, goffman_supercurrent_2000, van_dam_supercurrent_2006, pillet_andreev_2010, deacon_tunneling_2010, Basset2014}.
A fruitful way to describe transport through the device is to consider the formation 
of electronic bound states at the junction known as Andreev bound states.
In thermodynamic equilibrium, at low temperatures, and for short junctions
the current-phase relation is determined
mainly by the phase dependence of the lowest energy Andreev bound state.
A wealth of experimental and theoretical work has been devoted to investigate 
the current-phase dependence in Josephson junctions and leads, for instance,
to the prediction \cite{bulaevskii1977superconducting, rozhkov_josephson_2001, buzdin_proximity_2005, martin-rodero_josephson_2011} 
and the observation \cite{baselmans_reversing_1999, ryazanov_coupling_2001, kontos_josephson_2002, van_dam_supercurrent_2006, cleuziou_carbon_2006, Flensberg0Pi2007} of a change of sign of the current-phase relation, the so called 0-$\pi$ transition.
This can be induced by the presence of magnetic moments (magnetic impurities
or ferromagnetic layer) or in a non-magnetic material by the repulsive Coulomb
interaction at the quantum dot forming the junction, as is observed in carbon
nanotubes \cite{cleuziou_carbon_2006, Bouchiat2009, Wernsdorfer2012, pillet_tunneling_2013} 
or semiconducting nanowire \cite{van_dam_supercurrent_2006} Josephson junctions.
At the basis of this transition is the change of the parity of the junction. 
In superconductors electrons are paired, but if in the quantum dot forming 
the Josephson junction, Coulomb repulsion is sufficiently large, 
the ground state will accommodate only one electron. 
At lowest order in the tunnelling, the Josephson current is suppressed, 
and at the next (4th) order it changes sign \cite{rozhkov_josephson_2001}, since Cooper pairs are 
re-composed by tunnelling with reversed spins.

Only recently a direct detection of the excited Andreev bound states has been possible with a series
of experiments that probed the Josephson junction by resonating microwaves irradiation
\cite{bretheau_exciting_2013, bretheau_supercurrent_2013}.
These experiments pointed out the importance of the coupling to the electromagnetic (EM)
environment and in particular to the quasi-particles present in the superconducting leads.
It is an established experimental fact that the density of quasi-particles does not vanish exponentially 
with the temperature as predicted by the BCS theory, but remains finite, even at the lowest temperatures \cite{riste_millisecond_2013, olivares_dynamics_2014}.
Environment-assisted absorption of quasi-particles can modify the junction parity, 
since an unpaired electron can fall in the quantum-dot. 
This process has been considered very recently for junctions where Coulomb interaction is negligible \cite{olivares_dynamics_2014}.

In this paper, we investigate the effect of parity transitions induced by the quasi-particle absorption and 
emission in presence of Coulomb interaction.
We consider the limit for which the superconducting gap is the largest energy scale, also 
known as atomic limit. 
We obtain an exact Born-Markov description of the system coupled 
to the EM environment.
In this approximation, the $\pi$-phase is indicated by the occupation of an
odd-parity state with a vanishing of the supercurrent.
This allows to describe in a consistent way the $0$-$\pi$ transition by taking into 
account the relaxation processes that induce parity changes.
We find that the presence of quasi-particles can completely wash out the $0$-$\pi$ transition,
and invalidate the usual arguments based on the parity of the lowest energy state.
The quasi-particles are necessary to let the system relax to the lowest energy ground state,
but at the same time, they allow dynamic transitions between states, smoothening the transition.
We also considered the effect of the irradiation of a microwave signal on the gate of the quantum
dot that can be used to probe the state of the junction in both the $0$- and $\pi$- phase.

%
%
%
%
%
\begin{figure}[ht]
\centering
   \includegraphics[width=1.0\columnwidth]{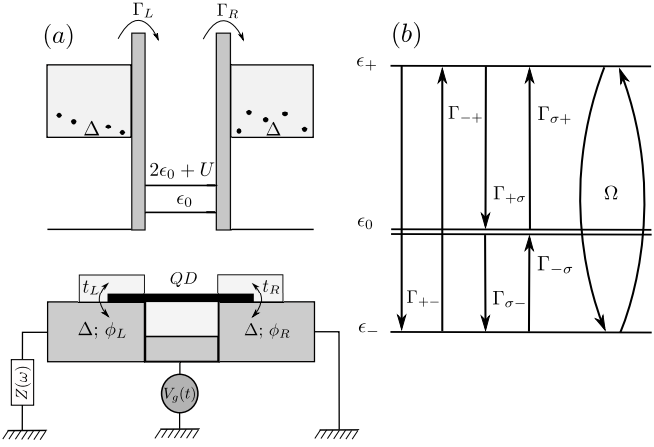}
\caption{\label{Fig0:fig} 
(a) Representation of the Josephson junction formed, for instance, by a carbon nanotube quantum dot
bridging two superconductors. (b) Schematics of the transitions between the Andreev bound states induced by incoherent
$\Gamma_{ij}$ and coherent $\Omega$ perturbations.}
\end{figure}
%
%
%
%

\textit{Model.---} 
Let us consider a quantum dot with a single electronic level  
forming a Josephson junction between two superconducting leads [see Fig. 1-(a)].
We assume that the junction is phase biased, that a time-dependent gate voltage can be applied,
and that the source-drain circuit is shunted on an impedance $Z(\omega)$.
This system can be modeled by the (time dependent) Hamiltonian
\begin{equation}
	\label{ham}
	H = H_{\rm dot}(t) + H_{\rm L} +H_{\rm R}+ H_{\rm T} + H_{\rm B} + H_{\rm c}
	\,.
\end{equation}
The first term of \Eqref{ham} reads $H_{\rm  dot}(t) = \epsilon_{d}(t)(n_\uparrow +n_\downarrow) 
+ U n_{\uparrow} n_{\downarrow}$ 
and it describes the single-electronic level of time-dependent 
energy $\epsilon_d(t)=\epsilon_0+\epsilon_1 \cos(\omega t)$ and Coulomb 
repulsion $U$. 
Here $n_\sigma=d_{\sigma}^{\dagger} d_{\sigma}$, and $d_\sigma$ is 
the electronic destruction operator on the dot of spin projection $\sigma$.
The second and third terms  
$H_{X} = \sum_{k\sigma} 
	\xi_{Xk} c_{Xk\sigma}^{\dagger}c^{\phantom \dag}_{Xk\sigma} +
	\sum_k 
		\left[
			\Delta_{X} e^{i\phi_{X}}c_{Xk\uparrow}^{\dagger}c_{X-k\downarrow}^{\dagger} + {\rm h.c.}
		\right]
$, describe the left and right leads ($X$=L,R) as BCS superconductors of order parameter  
$\Delta_X e^{i \phi_X}$ and electronic spectrum $\xi_{Xk}$, with $c_{Xk\sigma}$ the related destruction operator
for momentum $k$.
%
%
The dot and the leads are coupled by the tunneling term 
$H_{\rm T} =\sum_{Xk\sigma} t_{Xk} c_{Xk\sigma}^{\dagger}d^{\phantom{\dagger}}_{\sigma} +{\rm h.c.}$,
that gives rise to a rate $\Gamma_X= \pi \rho_X | t_X |^2/\hbar$, where $\rho_X$ is the 
density of states at the Fermi level of the superconductor $X$ and 
$\hbar$ the reduced Planck constant.
The EM modes described by $Z(\omega)$ induce 
fluctuations of the superconducting phase difference at the ends of the junction.
For simplicity we assume that the gate capacitance is much smaller than
the symmetric left and right capacitances. 
Within these assumptions and for ${\rm Re } (Z) \ll R_Q=\pi \hbar/2e^2$, the quantum of resistance 
($e$ electron's charge),
we can expand the dependence of the Hamiltonian on the
phase difference fluctuations $\tilde \phi$, obtaining the linearized coupling term:
$H_{\rm c}=(\hbar/e) I \tilde \phi$, where $I=(I_L-I_R)/2$ is the total physical current (including displacement current).
It is expressed in terms of the left and right particle current operators 
$I_X=i (e/\hbar) \sum_{k\sigma} t_{Xk} c_{Xk\sigma}^{\dagger}d^{\phantom{\dagger}}_{\sigma} -{\rm h.c.}$.
The term $H_B$ describes the EM modes and following Ref.~\cite{ingold_charge_1992} 
one obtains:
$
\langle \tilde \phi(t) \tilde \phi(0) \rangle \equiv \mathcal{C}_\phi(t) = 2 \int_0^\infty  
d \omega {\rm Re} [Z(\omega)]  [{\rm coth} (\hbar\omega/2 k_B T_{\rm EM}) \cos \omega t-i \sin \omega t]/(\omega  R_Q),
$
with $T_{\rm EM}$ the temperature of the EM environment and $k_B$ the Boltzmann constant. 
In the following we will consider the symmetric case, for which $\Gamma_X=\Gamma/2$, 
and $\Delta_X=\Delta$ for $X=$ L and R.
Moreover since the final results depend only on the phase difference $\phi_L-\phi_R$ we 
set from the outset $\phi_L=-\phi_R=\phi/2$.

\newcommand{\uda}{2}
\newcommand{\proj}[2]{|#1\rangle\langle#2|}

When the driving and the coupling to the environment is neglected this Hamiltonian has
been widely studied in the literature
\cite{glazman1989resonant,rozhkov_josephson_1999,vecino_josephson_2003,Martin0Pi2007,TMeng0Pi2009} and
it is known to show a rich phase diagram with 
a 0-$\pi$ transition controlled by Kondo correlations.
The problem can be treated analytically only in few regimes, and only for the equilibrium case 
an exact solution is available based on the numerical renormalization group 
\cite{YoshiokaNRG2000,choi_kondo_2004} or Monte Carlo simulations
\cite{QMCEgger2004}. 
The objective of this work is to explore the fate of the 0-$\pi$ transition in presence of the 
quasi-particles and EM environment.
The system being out-of equilibrium, we choose to investigate the case 
$\Delta \gg |\epsilon_d |, U, \hbar\Gamma, \hbar\omega$, for which a systematic controlled approximation is possible.
In this limit the four states of the isolated dot 
$\ket{0}$, 
$\ket{\uparrow}=d_{\uparrow}^{\dagger}\ket{0}$, 
$\ket{\downarrow}=d_{\downarrow}^{\dagger}\ket{0}$, and
$\ket{\uda}=d_{\uparrow}^{\dagger}d_{\downarrow}^{\dagger}\ket{0}$, 
are only weakly coupled to the leads, and their (unperturbed) energy levels 
$\{0$, $\epsilon_0$, $\epsilon_0$, $2\epsilon_0+U\}$ are well separated 
from the quasi-particle continuum.
Following a standard procedure of atomic physics \cite{Cohen-Tannoudji_Atom_Photon_Interaction} 
the effect of $H_T$ can then be taken into account systematically by performing a unitary transformation that 
generates an effective Hamiltonian $H_d^{\rm eff}$ in the 4-dimensional space of the quantum dot.
At lowest order in $\hbar\Gamma/\Delta$ one obtains
$H_d^{\rm eff}=
\epsilon_0( |\uparrow\rangle\langle\uparrow |+\proj{\downarrow}{\downarrow})
+(2\epsilon_0+U)\proj{\uda}{\uda}+\hbar\Gamma \cos (\phi) (|0\rangle \langle \uda|+| \uda \rangle \langle 0|)$,
where the last off-diagonal term hybridizing the even-parity states is a manifestation of the proximity effect. 
Performing the same unitary transformation on the current operator $I$ one obtains at the first two non-vanishing orders:
$I^{\rm eff}=I^{(1)}+I^{(2)}$, where 
$I^{(1)}=(e/\hbar)\sum_{\sigma,\alpha=\pm} D_{\alpha \sigma} C_{\alpha \sigma}$,
%
	$
	I^{(2)}=-e\Gamma\sin(\phi/2)(|0\rangle \langle \uda|+|\uda\rangle \langle 0|)$,
%
with
$D_{+\sigma}=\proj{\sigma}{0}+s_\sigma \proj{\uda}{\overline\sigma}$, $D_{+\sigma}=D_{-\sigma}^\dag$, 
$s_{\uparrow,\downarrow}=\pm 1$,
$
C_{\alpha\sigma} 
=
-i\alpha\sum_{X} (s_X t_{X}/2)\sum_{k} 
(u_{k}\gamma_{Xk\sigma}^{\bar{\alpha}} - s_\sigma v_{k}e^{-i\alpha\phi_{X}}\gamma_{X-k\bar{\sigma}}^{\alpha}) 
$.
The Bogoliubov operators $\gamma_{Xk\sigma}^{\alpha}$ diagonalize the BCS Hamiltonian of lead
$X$: $H_X=\sum_{k\sigma} E_k \gamma_{Xk\sigma}^{+}\gamma_{Xk\sigma}^{-}$,
with $\gamma_{Xk\sigma}^{+}$ and $\gamma_{Xk\sigma}^{-}$ indicating the creation and destruction operator 
for energy $E_{k} = (\xi_{k}^{2}+\Delta^{2})^{1/2}$. 
Finally $u_{k} (v_{k}) = [(1/2)(1 \pm \xi_{k}/E_{k})]^{1/2}$ and  $s_{L,R}=\pm1$.

\textit{Born-Markov description.---} 
In order to give a quantitative description of the dynamics
we proceed by treating the coupling to the environment by a Born-Markov approximation 
\cite{Cohen-Tannoudji_Atom_Photon_Interaction}.
We will regard the quasi-particles in the superconductor and the EM excitations 
as a Markovian environment. 
We describe the stationary distribution of quasi-particles as an equilibrium one
characterized by a temperature $T_{\rm qp} \gg T_{\rm EM}$, as it appears to be the case in 
several experiments \cite{bretheau_exciting_2013, bretheau_supercurrent_2013,olivares_dynamics_2014}.
Following the standard procedure and tracing out the quasi-particles and the EM fluctuations the 
equation for the reduced density matrix $\rho$ for the degrees of freedom of the dot reads:
\begin{eqnarray} 
\lefteqn{
\dot{\rho}(t) = -(i/\hbar)\lbrack H^{\rm eff}_{\rm d}(t),\rho(t) \rbrack  
+ (\hbar/e)^2\mathcal{L}_{\mathcal{C}_\phi} \lbrack I^{(2)}, I^{(2)} \rbrack}
\nonumber\\
&&+ 
\sum_{\alpha\sigma} 
\left\{
\mathcal{L}_{\mathcal{C}_N^{\alpha\sigma}} \lbrack D^{\phantom \dag}_{\alpha\sigma}, D^\dag_{\alpha\sigma} \rbrack   
+ 
\mathcal{L}_{\mathcal{C}_A^{\alpha\sigma}} \lbrack D^{\phantom \dag}_{\alpha\sigma}, D^{\phantom \dag}_{\alpha \bar{\sigma}} \rbrack    
\right\}\, .
\label{masterEq} 
\end{eqnarray}
Here
$
\hbar^2\mathcal{L}_{\mathcal{C}} \lbrack A, B \rbrack = -\int_{0}^{+\infty}d\tau
\left\{
	\mathcal{C}(\tau)\left[A, B(t-\tau,t) \rho(t) \right] 
	 \right. + \left. \mathcal{C}(-\tau)\left[\rho(t)A(t-\tau,t), B\right] 
\right\}
$, and we have defined the normal 
$\mathcal{C}^{\alpha\sigma} _{\rm N}(t)=\langle C^{\phantom \dag}_{\alpha\sigma}(t)C^{\dag}_{\alpha\sigma}(0)\rangle \mathcal{C}_\phi(t)$,
and anomalous
$\mathcal{C}^{\alpha\sigma} _{\rm A}(t)=\langle C^{\phantom \dag}_{\alpha\sigma}(t)C^{\phantom \dag}_{\alpha\bar{\sigma}}(0)\rangle \mathcal{C}_\phi(t)$
quasi-particles correlation function.
The second term of the right-hand side of \Eqref{masterEq} affects the evolution of only the even-parity states,
since $I^{(2)} $ has non-vanishing matrix elements only in this sub-space.
By contrast the third and fourth terms, that are generated by the $(\hbar/e) I^{(1)} \tilde\phi$ term,
allow a change in the parity of the dot. 
This is possible, since $I^{(1)}$ describes the transfer of one electron from (to) the leads to (from) 
the dot by the photon assisted destruction (creation) of a quasi-particle in the leads.
Within the approximations of the model one can obtain explicit expressions for the correlations functions.
We define $J(\omega)=2{\rm Re} [Z(\omega)]/(R_Q\omega)$, 
$
\mathcal{C}^{\alpha\sigma} _{\rm N,A}(\omega)
=
\int_\Delta^\infty  d E/\hbar
\left[J(E/\hbar+\omega)[1+n_B(E+\hbar\omega)]f_F(E) g_{1\rm N,A}^{\alpha \sigma}(E)
\right.
+
\left.
J(E/\hbar-\omega)n_B(E-\hbar\omega)[1-f_F(E)] g_{2\rm N,A}^{\alpha \sigma}(E)
\right]
$, with 
$n_B(E)=1/(e^{E/k_{\rm B} T_{\rm EM}}-1)$, 
$f_F(E)=1/(e^{E/k_{\rm B} T_{\rm qp}}+1)$, 
$g_{1(2)N}^{\alpha\sigma}(E) = 
(\Gamma/4)
[E/(E^2-\Delta^2)^{1/2}\mp \alpha]
$
and 
$g_{1(2)A}^{\alpha\sigma}(E) = 
\pm\alpha s_\sigma \Gamma \Delta \cos(\phi/2)/(4\sqrt{E^2-\Delta^2})$,
the master equation can now be solved numerically for a given choice 
of $Z(\omega)$ projecting on the Floquet basis \cite{grifoni_driven_1998}.
In the following we will discuss different regimes for which 
the analytical and numerical results will be compared.

{\em Non driven case.---}
When the driving term is absent $(\epsilon_1=0)$ the effective Hamiltonian can be 
easily diagonalized. 
The four states split into a degenerate doublet of
odd parity at energy $\epsilon_0$ and a non-degenerate pair of states generated by the 
hybridization of the even-parity states $|0\rangle$ and $|2\rangle$:
$\ket{-}=\cos(\beta)\ket{0}+\sin(\beta)\ket{2}$ and 
$\ket{+}=-\sin(\beta)\ket{0}+\cos(\beta)\ket{2}$.
Their energy reads 
$\epsilon_{\pm} = \epsilon_{0} + U/2 \pm [ (\hbar\Gamma\cos(\phi/2))^2+ (\epsilon_{0} + U/2)^{2}]^{1/2}$
with $\tan \beta=\epsilon_-/\hbar\Gamma \cos(\phi/2)$.
In this limit, and neglecting the environment, the transition 0-$\pi$ is particularly 
simple. 
Depending on the value of $U$ and $\epsilon_0$ the ground state can be either 
the even-parity state $|-\rangle$ (for $\epsilon_-<\epsilon_0$) 
or the two degenerate odd-parity states $|\sigma\rangle$ (for $\epsilon_->\epsilon_0$).
The current is simply obtained by the evaluation of the current operator on the ground state 
and it vanishes for the odd-parity states, while it equals 
$I_{--}=-e\Gamma \sin(2\beta)\sin(\phi/2)(=-I_{++})$ in 
the $|-\rangle$ state. 
It is known that going to the next order in the perturbation theory one obtains a negative 
value of the current in the odd-parity state \cite{rozhkov_josephson_2001}. 
In the following we will use the information on the parity of the occupied state to 
distinguish between the $0$ and the $\pi$ phase.
We can now discuss the effect of the environment, as predicted by \Eqref{masterEq}. 
In the absence of driving one can show that the density matrix becomes diagonal in 
the eigenstate basis of $H_{\rm d}^{\rm eff}$ and the effect of the environment reduces 
to a description of incoherent tunnelling between states. 
Neglecting the principal parts in \Eqref{masterEq} we obtain an explicit expression for the 
rates [see Fig. 1-(b)].
The transition inside the even-parity doublet is dominated by the direct coupling to the EM environment:
$\Gamma_{+-} = 2 \pi J(\delta_{+-})
[1 + n_{B}(\hbar\delta_{+-})](\hbar/e)^2 \left| I_{+-}\right|^2$,
$\Gamma_{-+} = 2\pi J(\delta_{+-})
n_{B}(\hbar\delta_{+-}) ({\hbar}/{e})^2|I_{+-}|^{2}$,
with 
$\hbar\delta_{+-}=\epsilon_{+} - \epsilon_{-}$,
$(\hbar/e)^2|I_{+-}|^{2}=\Gamma^2 \cos^2(2\beta) \sin^2(\phi/2)$.
The parity-breaking transitions $\Gamma_{a\sigma}$ with $a=\pm$
have the same form of $\mathcal{C}_{\rm N}^{\alpha \sigma}(\omega)$, 
with $\hbar\omega=\epsilon_a-\epsilon_\sigma$, and with 
$g_{1(2)N}^{\alpha\sigma}(E) \rightarrow 
(\Gamma/2)\lbrace [E\mp a \Delta \sin(2\beta)\cos(\phi/2)]/(E^2-\Delta^2)^{1/2}\mp a \cos(2 \beta)\rbrace
$.
The opposite transitions $\Gamma_{\sigma a}$ are obtained in the same way performing the substitution, 
$\hbar\omega \rightarrow \epsilon_\sigma-\epsilon_a$, and $g_{1(2)N}^{\alpha\sigma}(E) \rightarrow 
(\Gamma/2)\lbrace [E\pm a \Delta \sin(2\beta)\cos(\phi/2)]/(E^2-\Delta^2)^{1/2}\pm a \cos(2 \beta)\rbrace
$. Assuming $k_BT_{EM}\ll \Delta$ and  ${\rm Re}[Z(\omega)] = \gammaEM \omega^2 $
for $\omega\lesssim \Delta$\footnote{We estimated the low
energy behavior of ${\rm Re}[Z(\omega)] \approx (2 L^2/RR_Q) \omega^2 $ 
from Refs.\cite{Desposito2001,bretheau_exciting_2013, bretheau_supercurrent_2013}, involving a typical 
Josephson junction inductance $L\approx 0.6 nH$, circuit resistance $R\approx 200 \Omega$ and 
superconducting gap $\Delta \approx 0.1 meV$.}, we approximate  
$J(\omega)=\gammaEM \omega$.
The expressions for the rates can then be further simplified 
performing the integrals in $E$. 
One obtains:
\begin{eqnarray} 
	\Gamma_{+-} &=& 2\pi\gammaEM \delta_{+-} \Gamma^2 \cos^2(2\beta)\sin^2({\phi}/{2}), 
	\nonumber
	\\
	\Gamma_{a\sigma/\sigma a} 
	&=& 
	 \gammaEM \frac{\Delta^2}{\hbar^2} \Gamma \Xi\left( \frac{k_B T_{\rm qp}}{\Delta} \right)
	  (1 \mp a u) [1 \pm (\epsilon_a-\epsilon_0)/\Delta], \nonumber
\end{eqnarray}  
with $\Gamma_{-+}=0$, 
$\Xi(x)=e^{-1/x} \sqrt{\pi x/8}$, and
$u=\sin 2\beta \cos(\phi/2)$.
(Note that $\beta$ depends on $\phi$ and the expressions for the rates are 
correctly $2\pi$ periodic in $\phi$.)
According to our approximation the energy dependence of the rate is very
weak, since $|\epsilon_a-\epsilon_0|\ll \Delta$.
This implies that the energy ordering of the two states has very little effect 
on the parity-breaking rates. 
The reason is clear: the transition from one state to the other is possible thanks to 
a quasi-particle of energy $\Delta$, that has to be present in the environment.
An electron can then be added or removed from the dot, and the excess energy 
is absorbed by a phonon. 
The relative energy of the initial and final states of the dot multiplet is small
with respect to $\Delta$, and thus in the end the energy ordering will not be 
important. 
In other terms the coupling to the environment will not allow a relaxation 
of the dot to its lowest energy state, but induce instead  
transitions from the 0 to the $\pi$ states.
The average measured current becomes then simply 
$I_{--}\rho_{--}$: The magnetic states do not carry current,
and the state $|+\rangle$ relaxes very rapidly to the 
state $|-\rangle$, since this transition does not need the participation of 
the rare quasi-particles. 
The final result is that the $0$-$\pi$ transition can be completely washed 
out in the average current.
This is is clearly visible in Fig. 2, where 
the numerical and analytical solution of \Eqref{masterEq} 
as a function of $U$ for $\phi=\pi/2$ is compared to the prediction of the system 
not coupled to the environment. 
The former has a smooth behavior following the $U$-dependence of $I_{--}$,
while the latter has a sharp jump.
A similar picture is obtained as a function of $\epsilon_0/\Gamma$.
%
%
%
%
%
\begin{figure}[ht]
\centering
   \includegraphics[width=1.0\columnwidth]{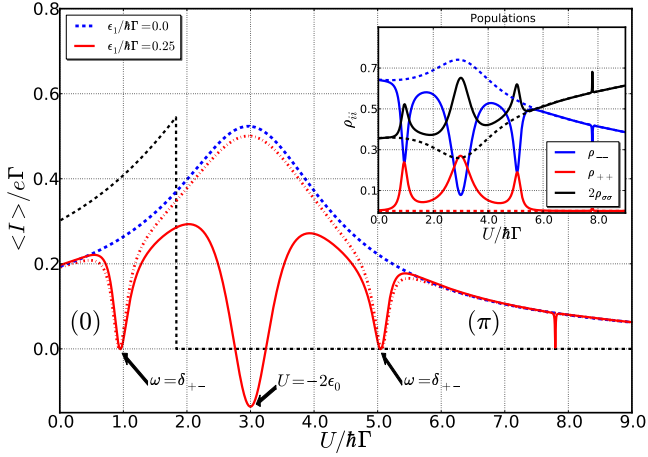}
\caption{\label{Fig2:fig} 
(Color online) Average current $\langle I \rangle$ as a function of Coulomb repulsion $U$. 
The black dashed curve results from 
the thermodynamic arguments in absence of driving. The full red (dashed blue) curve is  
the numerical solution of the Floquet master equation in presence (absence) of AC driving. 
The red dash-dotted curve is the outcome of the optical Bloch equations in the RWA approximation. 
Inset: The corresponding 
plain (dashed) numerical curves for the populations $\rho_{++},\rho_{--}$ and $2\rho_{\sigma\sigma}$ of the Andreev 
bound states in presence (absence) of AC driving. The parameters describing the 
Josephson junction are common to both plots: $\Delta/\hbar\Gamma=10.0, \epsilon_{0}/\hbar\Gamma=-1.5, \epsilon_{1}/\hbar\Gamma=0.25, \phi=\pi/2, \omega/\Gamma=2.5, k_{B}T_{EM}/\Delta=0, k_{B}T_{qp}/\Delta=1/20$ and $\gammaEM\Gamma^2=1.4 \times 10^{-4}$.}
\end{figure}
%
%

{\em Effect of driving.---}
An experimental way of testing the state of the junction is to irradiate the gate with an AC field. 
The resulting modulation in time of $\epsilon_d(t)$ is a perturbation that cannot change the parity 
of the junction.
Since the odd-parity states are degenerate, the AC field can only induce resonant transitions between 
the even-parity states for small values of the detuning $\delta=\omega-\delta_{+-}$.
Close to the resonance the dynamics can be described by performing a rotating-wave approximation
that gives for the density matrix in the rotating frame $\tilde{\rho}_{a\bar{a}}=\rho_{a\bar{a}}e^{ia\omega t}$,
$\tilde{\rho}_{aa}=\rho_{aa}$:
\begin{eqnarray} 
\dot{\tilde{\rho}}_{aa} 
&=& -i\frac{\Omega}{2} \lbrack \tilde{\rho}_{\bar{a}a} - \tilde{\rho}_{a\bar{a}} 
\rbrack\\
&-& 
(\Gamma_{a\bar{a}} + 2\Gamma_{a\sigma})\tilde{\rho}_{aa}
+ \Gamma_{\bar{a}a}\tilde{\rho}_{\bar{a}\bar{a}}
+ 2\Gamma_{\sigma a}\tilde{\rho}_{\sigma\sigma} , 
\nonumber \\
\dot{\tilde{\rho}}_{a\bar{a}} &=& -i\frac{\Omega}{2}
\lbrack \tilde{\rho}_{\bar{a}\bar{a}} - \tilde{\rho}_{S;aa}
 \rbrack \label{eqn:OpticalBloch2} \\ 
&+& 
	[	
	2ia\delta -
	(\Gamma_{a\bar{a}}+\Gamma_{\bar{a}a}+2\Gamma_{a\sigma}+2\Gamma_{\bar{a}\sigma})
	]
	\tilde{\rho}_{a\bar{a}}/2 ,
\nonumber 
\end{eqnarray}  
with
$\tilde{\rho}_{\sigma\sigma} = (1-\tilde{\rho}_{++}-\tilde{\rho}_{--} )/2$, $\tilde{\rho}_{\uparrow\downarrow}=\tilde{\rho}_{\downarrow\uparrow}\approx 0$, $\tilde{\rho}_{\bar{a}a} = \tilde{\rho}^{*}_{a\bar{a}}$
and $\hbar\Omega=\epsilon_1\sin(2\beta)$.
As in the optical Bloch equations, when $\Omega\gg \Gamma_{+-},\Gamma_{a\sigma}$ the coherence 
terms are important for the time evolution of the system.
The rates implying the quasi-particles are much smaller than all the other quantities
appearing in the master equation. 
Using this fact one can solve the equations for the block $+/-$ 
for given $\rho_{\sigma\sigma}(t)$ and then solve separately the resulting 
equation for $\rho_{\sigma \sigma}$.
This gives at vanishing $T_{EM}$:
\begin{equation}
	\rho_{++}=
		{
		\left[1+2\Gamma_{-\sigma}/\sum_a \Gamma_{\sigma a}\right]^{-1} 
			(\Omega/2)^2 
			\over 
		\delta^2+\left(\Gamma_{+-}^2+\Omega^2(1+\theta) \right)/4
		}
			\label{rhoPP}
\end{equation}
with $\theta=(2 \Gamma_{+\sigma}+\sum_a \Gamma_{\sigma a})/(2 \Gamma_{-\sigma}+\sum_a \Gamma_{\sigma a})$,
and 
$\rho_{++}/\rho_{--} = (\Omega/2)^2/[\delta^2+\left(\Gamma_{+-}^2+\Omega^2\right)/4]$.
Eq.\refe{rhoPP} describes a typical resonant behavior for the populations as it can be seen 
in the inset of Fig. 2.
At resonance ($\delta=0$) the populations equilibrate so that 
$\rho_{++}=\rho_{--}=1-2\rho_{\sigma\sigma} =
\sum_a \Gamma_{\sigma a} / 2\sum_a(\Gamma_{a\sigma}+\Gamma_{\sigma a})$.
Since typically the rates are of the same order of magnitude at resonance 
$\rho_{++} \approx 1/4$. 
The average current $\langle I \rangle$ is simply $\sum_a \rho_{aa} I_{aa}$ and it is 
strongly modulated near the resonances.
The resonance is visible in both the regions where the 0- and $\pi$ phase
would be stable. 
%
%
The narrow dip in Fig. 2 for $U/\hbar\Gamma\approx 8.0$ is a two-photon resonance  
described by the full numerical solution of Eq.\refe{masterEq}.  
%

%
%

The slow fluctuations between the $\sigma$ and $\pm$ doublets induce a 
strong telegraph noise, since the current in the four states is very different
and the fluctuations are slow. 
To estimate the intensity of the current noise we assume that all current fluctuations
are due to the transitions among the four states, each one having a different value for 
the stationary current (specifically $I_{++}=-I_{--}$, and $I_{\sigma\sigma}=0$).
In absence of driving one finds that the current noise reads:
$S=4\Gamma_{\sigma-}\Gamma_{-\sigma}I_{--}^2/(\Gamma_{\sigma-}+2 \Gamma_{-\sigma})^3$,
giving a very large Fano factor $F=S/2eI$ of the order of $e^{\Delta/k_{B}T_{\rm qp}}$.
We note that a strong telegraph noise has been very recently observed in atomic point 
contact junctions \cite{bretheau_supercurrent_2013}. In that experiment the Coulomb blockade plays no role, but 
the coupling to the quasi-particles has a very similar behavior. 
%
%

The driving reveals also an unexpected maximum of the population $\rho_{++}$ for $U=3\Gamma$. 
This is the value for which $2\epsilon_0+U=0$ and the $\ket{0}$ and $\ket{2}$ states are degenerate. 
At this point the matrix element entering the rate $\Gamma_{+-}$ vanishes.
The population of the excited state generated by the non-resonant driving can relax
to the $\ket{-}$ state only passing through the $\ket{\sigma}$ states, with very low rates.
This allow a large population of the excited state with a consequent negative contribution 
to the supercurrent.

%
%

\textit{Conclusion.---} 
We have investigated the effect of a coupling to the quasi-particles and the
EM environment on the 0-$\pi$ transition.
In a regime where the approximations can be well controlled we have shown that
the quasi-particle scattering induces transitions between the
$0$- and $\pi$- states, with a consequent washing out of the transition.
We found that this induces large current fluctuations, and that the state of
the junction could be investigated by driving the gate with an AC voltage.
The main reason for the smoothening of the transition is the fact that the excess energy
of the quasi-particles allows fluctuations from the thermodynamical ground state
and the first excited state.
This effect is very strong in the regime where we work, since $\Delta$ is the
largest energy scale, but it will be present also for intermediate values
of the gap.
The theory we present indicates clearly that the effect of quasi-particles can be dramatic.
The question of the crossover to the thermodynamical equilibrium when $\Delta$ is
of the same order or smaller than the other energy scales remains open and
calls for further investigations.
The issue of the stability of Andreev bound states with respect to the
quasi-particle scattering has also a strong relevance for the observation of Majorana
states, that should be subject to a similar dynamics \cite{Kane2009}.


We acknowledge financial support from the ANR QNM $n^{\circ}$ 0404 01 and the
PHC NANO ESPAGNE 2013 project $n^{\circ}$ 31404NA.
Useful discussions with A. Levy Yeyati and D. G. Olivares are acknowledged. 
We thank for comments M. Houzet and M. F. Goffman.

\bibliography{Superconductor_Relaxation_Paper}

\end{document}